\documentclass[journal]{IEEEtran}
\usepackage{colortbl}
\usepackage[caption=false,font=normalsize,labelfont=sf,textfont=sf]{subfig}
\usepackage{caption}
\usepackage[most]{tcolorbox}
\usepackage{booktabs}
\usepackage{cite}
\usepackage{xspace}
\usepackage{makecell}
\usepackage{listings}
\usepackage{hyperref}
\tcbset{
    colback=gray!10,
    colframe=black,
    listing only,
    listing options={basicstyle=\ttfamily\scriptsize, language=C, breaklines=true},
    left=0mm, right=0mm, top=0mm, bottom=0mm,
    boxrule=0.3mm,
    width=\linewidth,
    halign=center
}
\hypersetup{
    colorlinks=true,
    linkcolor=blue,     
    citecolor=blue,      
    filecolor=blue,     
    urlcolor=blue     
}
\usepackage[frozencache,cachedir=.]{minted}
\usepackage{pifont}
\usepackage{array}
\newcommand{\cmark}{\ding{51}}

\newcolumntype{C}[1]{>{\centering\arraybackslash}p{#1}}

\newcommand{\esbmcai}{{\sc ESBMC-AI}\xspace}

\begin{document}

\title{A New Era in Software Security: Towards Self-Healing Software via Large Language Models and Formal Verification}

\author{
\IEEEauthorblockN{Norbert Tihanyi, Ridhi Jain, Mohamed Amine Ferrag}
\IEEEauthorblockA{(Technology Innovation Institute (TII), UAE)\\
}
\and
\IEEEauthorblockN{ Yiannis Charalambous, Youcheng Sun,  Lucas C. Cordeiro}
\IEEEauthorblockA{(University of Manchester, UK)\\}
}

\markboth{}%
{Shell \MakeLowercase{\textit{et al.}}: A Sample Article Using IEEEtran.cls for IEEE Journals}

\maketitle

\begin{abstract}
This paper introduces an innovative approach that combines Large Language Models (LLMs) with Formal Verification strategies for automatic software vulnerability repair. Initially, we employ Bounded Model Checking (BMC) to identify vulnerabilities and extract counterexamples. These counterexamples are supported by mathematical proofs and the stack trace of the vulnerabilities. Using a specially designed prompt, we combine the original source code with the identified vulnerability, including its stack trace and counterexample that specifies the line number and error type. This combined information is then fed into an LLM, which is instructed to attempt to fix the code. The new code is subsequently verified again using BMC to ensure the fix succeeded. We present the \esbmcai framework as a proof of concept, leveraging the well-recognized and industry-adopted Efficient SMT-based Context-Bounded Model Checker (ESBMC) and a pre-trained transformer model to detect and fix errors in C programs, particularly in critical software components. We evaluated our approach on $50,000$ C programs randomly selected from the FormAI dataset with their respective vulnerability classifications. Our results demonstrate \esbmcai's capability to automate the detection and repair of issues such as buffer overflow, arithmetic overflow, and pointer dereference failures with high accuracy. \esbmcai is a pioneering initiative, integrating LLMs with BMC techniques, offering potential integration into the continuous integration and deployment (CI/CD) process within the software development lifecycle.
\end{abstract}


\section{Introduction}
\label{sec:introduction}

Implementation bugs can impact the software quality by causing crashes, data loss, poor performance, or incorrect results~\cite{zaman2011security, zhang2010conmem}. These bugs often lead to vulnerabilities, emphasizing the need for early identification and resolution~\cite{bajwa2015unintentional}. Consequently, automated software testing~\cite{dustin1999automated, godefroid2008automating, AldughaimAGFC23}, fault localization~\cite{AlvesCF17}, and repair~\cite{goues2019automatedrepair} have been active research areas over the past few decades. While classic static analysis aids early bug detection, it introduces false positives impacting developer productivity~\cite{sadowski2014developers, GadelhaSC0N19}. Recent deep learning (DL) advancements have drawn the attention of the software engineering community, offering potential solutions to longstanding issues~\cite{white2016deep, zhao2018deepsim, gupta2017deepfix}. For example, DLFix~\cite{li2020dlfix} and DeepRepair~\cite{deeprepair} treat source code as text; however, as opposed to natural language, source code has a stronger syntax and semantics~\cite{cure}; further, as these approaches rely on previously seen data, they may produce incorrect results. Often, this 
comprises small snippets of buggy code~\cite{li2020dlfix, chen2019sequencer, lutellier2020coconut}; thus, the model may not have the details of the bug, its origin, and how it interacts with the rest of the program. Contrarily, CURE~\cite{cure} employs a programming language model to parse, analyze, and model the source code. DEAR~\cite{li2022dear} combines spectrum-based fault localization with DL to learn the appropriate code-context.

\begin{figure*}[ht] 
    \centering
    \includegraphics[width=0.95\textwidth]{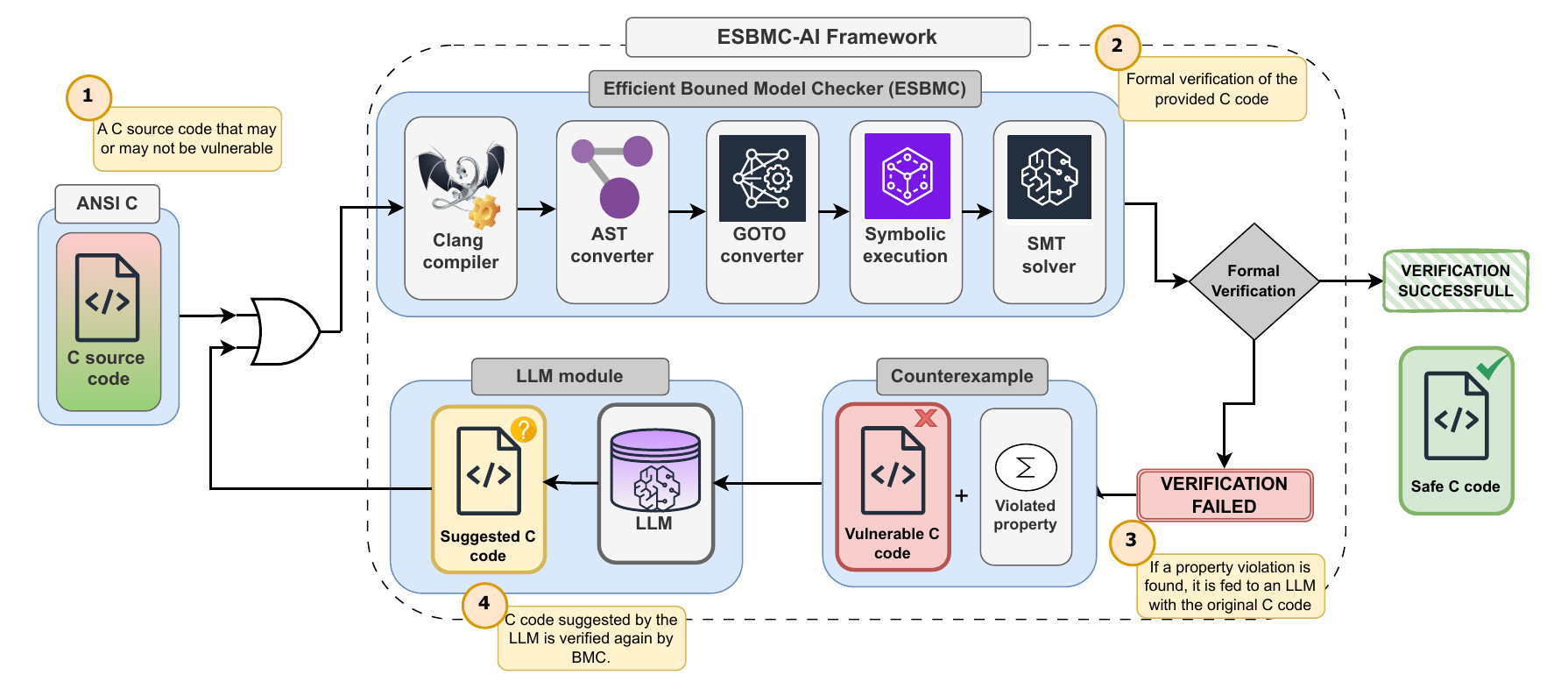}
    \setlength{\belowcaptionskip}{-10pt}
 \caption{An overview of the \esbmcai framework. Initially, a C source code is verified with ESBMC. If the verification fails, the property violation output from ESBMC, along with the original C code, is fed to the LLM to obtain the potentially fixed code. This process is repeated for the generated C code until it can be successfully verified by ESBMC.}
    \label{bmc-llm-approach}
\end{figure*}

Recent advances in Large Language Models (LLMs) such as OpenAI's Codex~\cite{chen2021evaluating}, a GPT-like LLM tailored for code program repair~\cite{prenner2021automatic, fan2022improving}, have demonstrated significant promise in addressing software engineering and testing challenges. For instance, InferFix~\cite{jin2023inferfix} applies LLMs to fix issues such as Null Pointer Dereference (NPD), Resource Leak (RL), and Thread Safety Violation (TSV). Xia et al.~\cite{xia2022practical} show that applying state-of-the-art LLMs directly can outperform existing automated program repair techniques. Indeed, leveraging LLMs holds potential in vulnerability detection and Automatic Code Repair (ACR)~\cite{pearce2022examining, cao2023prompt, xia2023automated}. However, deploying LLMs in software verification has limitations. Notably, state-of-the-art LLMs struggled to respond accurately when verifying software containing arithmetic expressions involving non-deterministic variables. In ACR, addressing a specific bug requires bit-precise calculations, appropriate Satisfiability Modulo Theory (SMT) solving skills, accurate parsing of the Abstract Syntax Tree (AST)~\cite{xin2011programAST}, and precise data flow analysis~\cite{sui2014detectingdfa}. These tasks demand precise and strict answers, where the non-deterministic behavior of LLMs can be problematic. Here, a precise external guide is needed for LLMs to pinpoint the exact location of vulnerabilities in the code. To address the unreliability of LLMs when used as stand-alone vulnerability detection tools, we propose integrating an LLM with the Efficient SMT-based Context-Bounded Model Checker (ESBMC)~\cite{gadelha2018esbmc} a well-recognized and industry-adopted Formal Verification (FV) tool, which produces very low false negative and false positive findings, thereby enhancing the method's efficiency. Figure~\ref{bmc-llm-approach} illustrates our counterexample-guided ACR methodology, combining BMC and LLM. The process involves the following steps: 
\ding{192} \textit{Initial Verification:} The BMC module takes the source code provided by the user and verifies or falsifies a property specification. \ding{193} \textit{Failure Handling:} If the verification fails, the BMC engine refutes the safety/security property. The original code and the counterexample for the property violation generated by BMC are then passed to the LLM module. \ding{194} \textit{Iterative Correction:} The LLM engine receives customized queries to produce potentially corrected code, which is then fed back to the BMC module to verify whether the corrected version meets the initial safety specification.

In this paper, we aim to address the following research questions:  

\begin{tcolorbox}[colback=gray!10]

\begin{itemize}
\item {\textbf{RQ1}:} Can we enhance the ACR capabilities of current LLMs by combining them with an external FV tool?

\end{itemize}

\end{tcolorbox}

\begin{tcolorbox}[colback=gray!10]

\begin{itemize}
\item {\textbf{RQ2}:} Which vulnerabilities are the most challenging for LLMs to repair successfully?

\end{itemize}

\end{tcolorbox}
\begin{tcolorbox}[colback=gray!10]

\begin{itemize}

\item {\textbf{RQ3}:} How reliable is the generated patch, and how likely is it that the new code remains compilable and does not disrupt the original program workflow?

\end{itemize}

\end{tcolorbox}

This research aims to study and identify the impact of formal verification tool-based feedback on LLMs' ability to repair faulty C code. The main original contributions of this work are as follows:
\begin{enumerate}
    \item We propose a novel software verification and repair approach, \esbmcai, which leverages the Efficient SMT-based Context-Bounded Model Checker (ESBMC) to provide stack traces and counterexamples of given vulnerabilities to an LLM for code repair;
    \item A comprehensive experimental assessment on over $1,000$ C programs, randomly selected from the FormAI dataset~\cite{tihanyi2023formai}, to examine the effectiveness of \esbmcai in repairing program codes; 

    \item Overall, formal program verification is undecidable~\cite{turing1936computable,sipser2012introduction}, making it impossible to create a computational method that can determine whether any given program is completely error-free. It is also ambitious to expect proof that a fixed patch does not disrupt the original program's functionality. To address this issue, we have calculated the cyclomatic complexity of each generated patch to identify any significant deviations from the original program. Additionally, each patch has been verified by at least one human expert to ensure that the accuracies reported in this study are as precise as possible.
    \item With \esbmcai, we achieved a code repair accuracy of $90.40$\% for buffer overflow on \texttt{scanf}, $86.47$\% for division by zero, $70.27$\% for arithmetic overflow on add, and $69.66$\% for array bounds violation errors. 
    \item We release \esbmcai 0.5.1\footnote{\url{https://pypi.org/project/esbmc-ai/}} as a PyPI module for industrial partners and the research community to use for ACR. Further details and results are available on the project website: \url{https://github.com/esbmc/esbmc-ai}
\end{enumerate}

This paper is organized as follows: Section~\ref{sec:motivation} provides motivating examples for this counterexample-guided code repair framework. We discuss prior related work in Section~\ref{sec:related}, while the background is discussed in Section~\ref{sec:preliminaries}. We present our experimental results in Section~\ref{sec:experiments}, which includes the experimental setup details and findings. Lastly, we conclude our work with future research directions in Section~\ref{sec:conclusion}. 

\section{Motivation Example}
\label{sec:motivation}

Detecting software vulnerabilities with LLMs is challenging, given their tendency to generate multiple alternative solutions for the same problem without explicitly identifying the root cause. While this adaptability is advantageous in language processing and text generation, it introduces complexities when proposing solutions for even trivial software security vulnerabilities~\cite{cao2023prompt, llmAPRuntrust, strasser2023pitfalls}. Although Bounded Model Checking (BMC) excels in identifying vulnerabilities through mathematical proofs, rectifying code based on BMC output demands a deep understanding of the tools and a background in formal verification. Automating code repair using BMC holds great potential for streamlining secure software development but has complexities. Despite their current limitations in software vulnerability identification, LLMs can significantly enhance their effectiveness by integrating the output from a BMC tool. This integration could lead to a surprisingly effective vulnerability detection and code repair approach. For example, consider the simple C code fragment in Listing~\ref{lis:arithmetic_overflow_ex}, where the initialization of variable $z$ triggers an arithmetic overflow.

\begin{listing}[H]
\caption{An arithmetic overflow example C program}
\begin{tcolorbox}[title={C++ program example}, coltitle=white, colbacktitle=black]
\scriptsize
\begin{minted}[xleftmargin=6pt,linenos,escapeinside=||]{c}
int main() {
    int x=77; 
    int y=x*x*x;
    int z=y*y;
    unsigned int r= z/1000;
    return 0;} 
\end{minted}
\end{tcolorbox}
\label{lis:arithmetic_overflow_ex}
\end{listing}

Upon querying GPT-3.5 about potential issues with the code fragment in Listing~\ref{lis:arithmetic_overflow_ex}, it responded: \textit{"There is no problem with this code. The code correctly calculates the value of $z$ as the square of $y$ and then calculates the value of $r$ as $z$ divided by 1000}. Even after rephrasing the question as \textit{``Is there any overflow in the program?''}, the model continued to respond inaccurately, specifically: \textit{"The value of $y$ is calculated as the cube of $x$, which is $77\times77\times77 = 456533$, and the value of $z$ is calculated as the square of $y$, which is $456533\times456533 = 208598234889$. This value can be represented within the $32$-bit int variable range, so there is no integer overflow."}

According to Imani et al.~\cite{imani2023mathprompter}, GPT models still struggle with evaluating arithmetic operations in programming languages. This is evident in the example where the model correctly calculates $456,533 \times 456,533 = 208,598,234,889$ but incorrectly claims that this value can be represented by a $32$-bit int variable, overlooking the maximum value a 32-bit integer can fit, i.e., $2^{32}=4,294,967,296$.

While the solution might be correct for this specific case when using more advanced LLMs such as the latest GPT-4o~\cite{openai2024-4o} or Falcon-180B~\cite{almazrouei2023falcon}, the same issue could arise when handling more complex arithmetic operations during program repair, especially when the BMC module generates a very complex SMT formula to be solved. Four claims must be resolved during verification, even for a small program, listed in Listing~\ref{lis:arithmetic_overflow_ex}. The SMT formula generated involves 64-bit BitVec operations, requiring a robust SMT solver such as Z3 ~\cite{de2008z3}, Boolector~\cite{brummayer2009boolector}, or CVC5~\cite{barbosa2022cvc5} to solve the underlying equation.

Initial experiments suggest that solving such straightforward examples with high confidence remains challenging. However, augmenting the capabilities of an LLM with an external tool can greatly improve its code repair abilities. Using BMC, we can easily detect the arithmetic overflow in our simple program:
\texttt{PROPERTY VIOLATION: line 4 function main arithmetic overflow on mul !overflow("*", y, y)}. When we ask the LLM to correct the code based on the specified property violation, supported by the stack trace and exact counterexample, we receive an accurate answer: \textit{``The issue is an arithmetic overflow on the line int $z=y*y;$. The int data type in C has a limited range of values that it can represent, typically $-2,147,483,648$ to $2,147,483,647$. 
In the line int $z = y*y$; there is a risk of integer overflow if the value of y is large. This large value exceeds an int data type's range, resulting in an arithmetic overflow.''}
We advocate that assistance from an external BMC tool can guide the LLM towards the correct solution. The LLM utilizes this to generate code that is safeguarded against arithmetic issues by substituting the $int$ data type with $long$ $long$ $int$, as illustrated in Listing~\ref{lis:arithmetic_overflow_fixed}.
\begin{listing}[H]
\caption{Corrected code for the code shown in Listing~\ref{lis:arithmetic_overflow_ex}}
\begin{tcolorbox}[title={Corrected C++ program suggested by the LLM}, coltitle=white, colbacktitle=black]
\scriptsize
\begin{minted}[xleftmargin=6pt,linenos,escapeinside=||]{c}
int main() {
    int x = 77;
    long long int y = (long long int)
    x * x * x;
    long long int z = y * y;
    unsigned int r = z / 1000;
    return 0;}
\end{minted}
\end{tcolorbox}
\label{lis:arithmetic_overflow_fixed}
\end{listing}

Upon running the BMC tool against the updated code, we received a \texttt{``VERIFICATION SUCCESSFUL''} output, indicating no integer boundary violations or overflows in the modified code. This small example provides convincing evidence that this approach is feasible and highly useful for ACR in industries requiring formal verification for critical software components. This motivates us to further investigate and explore this promising research direction in greater detail.

\section{Related Work}
\label{sec:related}

\subsection{Traditional Vulnerability Detection}

Traditional vulnerability detection methods often rely on static~\cite{gao2018static, medeiros2016dekantStatic, liang2016mlsaStatic} and dynamic~\cite{jain2022bird, chen2020muzzDynamic, jeong2019razzer} analysis techniques to identify security weaknesses in software. Although static approaches, including static code analysis~\cite{sadowski2015tricorder, aftandilian2012building, calcagno2015moving}, abstract syntax tree (AST) parsing~\cite{ma2020rejection, xin2011programAST}, and data flow analysis~\cite{sui2014detectingdfa, sampaio2016exploringdfa} enable early detection, they have high false positive rates~\cite{guo2023mitigating}. In contrast, dynamic analysis techniques, such as penetration testing~\cite{lee2020fusepen, li2015potassiumpen, salas2015blackpen}, fuzz testing~\cite{chen2020muzzDynamic, jeong2019razzer}, and runtime monitoring~\cite{arnold2011qvm, varvaressos2017automated, asadollah2018runtime}, provide a more realistic assessment by evaluating software behavior during execution. However, these approaches are often input-dependent, provide only partial code coverage, and are expensive. Hybrid approaches~\cite{smaragdakis2007hybrid, bhayat2021towards, aljaafari2022combining} combine static and dynamic analysis to balance their strengths and weaknesses. Bhayat et al.~\cite{bhayat2021towards} propose a comprehensive strategy integrating pre- and post-deployment techniques. Pre-deployment involves identifying vulnerabilities through static analysis using bounded model checking and symbolic execution. Post-deployment focuses on mitigating these vulnerabilities through hardware measures and software runtime protection. The hybrid approach underscores the effectiveness of integrated protection over individual components. Aljaafari et al.~\cite{aljaafari2022combining} proposed Ensembles of Bounded Model Checking with Fuzzing (EBF) that combine BMC with Gray-Box Fuzzing (GBF) in OpenGBF to detect software vulnerabilities in concurrent programs. 

Alternately, BMC provides reliable results with reduced costs as they limit the exploration depth for the test program. Song et al.~\cite{song2022esbmc} introduce ESBMC-Solidity, a Solidity frontend for ESBMC designed to verify the security of smart contracts on Ethereum's blockchain network. Alshmrany et al. \cite{alshmrany2022fusebmc} present an upgraded version of FuSeBMC, a tool that uses BMC and Evolutionary Fuzzing engines for improved code coverage and bug detection.
However, these approaches do not scale well even with the restricted depth exploration.

\subsection{Deep Learning-based Vulnerability Detection}

DeepFix~\cite{gupta2017deepfix} is a multi-layer sequence-to-sequence neural network that can fix compile-time errors. S{\footnotesize EQUENCE}R~\cite{chen2019sequencer} employs a similar technique to fix logical bugs by suggesting single-line patches, requiring a larger vocabulary. VRepair generates multiline patches using transfer learning~\cite{vrepair}. GetaFix~\cite{getafix} learns to generate patches by analyzing past human commits. Similarly, DEAR~\cite{li2022dear} uses AST-differencing to learn fine-grained changes and implements fault localization to identify problematic statements and produce relevant patches. DEAR and several other studies~\cite{zhu2021syntaxnmt, lutellier2020coconut} model ACR as a Neural Machine Translation (NMT)~\cite{sutskever2014nmt} problem. DeepRepair~\cite{deeprepair} uses DL code similarity to generate and validate patches. Huang et al.~\cite{huang2023empirical} leverage Large Language Models of Code (LLMCs) for  ACR by fine-tuning these models under the NMT paradigm.

Latest advancements in DL, transformers, and LLMs have revolutionized natural language processing, enabling machines to understand and generate human-like language~\cite{rajasekharan2023reliable, ge2023openagi}. These models can process vast amounts of textual data and extract meaningful information, making them useful tools for applications such as language translation, text summarization, sentiment analysis, and question-answering systems. LLMs' ability to generate code~\cite{huang2023agentcoder, chakraborty2022natgen, dakhel2023github} has made them a popular candidate for ACR~\cite{alpharepair, jin2023inferfix, dakhel2023github, jain2022jigsaw}. 

Many studies on LLM for ACR evaluate their approaches~\cite{sobania2023analysis, cao2023prompt} on QuixBugs~\cite{lin2017quixbugs}, containing only Java and Python test programs. Researchers have also investigated the potency of GPT in identifying and repairing software bugs~\cite{lajko2022fine, lajko2022towards, sobania2023analysis, pearce2022examining, cao2023prompt, xia2023automated}. Self-Edit~\cite{zhang2023selfedit} employs a generate-and-edit approach using test execution results from LLM-generated code to fix and improve code quality. RepairAgent~\cite{bouzenia2024repairagent} is an LLM-based agent for program repair, enabling dynamic bug-fixing through interaction with bug information, repair tools, and validation mechanisms. SecRepair~\cite{islam2024llm}, leveraging CodeGen2 and reinforcement learning, identifies and fixes vulnerabilities with descriptive code comments. MOREPAIR~\cite{yang2024multiobjective} introduces a fine-tuning approach for LLMs in ACR, emphasizing syntactic adaptation and logical reasoning behind code changes. 

 GPT models, with billions of parameters, produce accurate and contextually aware language models that are customizable through fine-tuning for specific tasks. Nonetheless, studies show that the codes and patches synthesized by GPT models may be incorrect and untrustworthy~\cite{llmAPRuntrust, evalplus, pearce2022examining, tian2023assistant, ma2023scope, liu2024codeevaluation}. New research proposes a prompt-based approach to verify the generated programs~\cite{xia2023conversational, compcode, hidvegi2024cigar}. The quality of fixes generated depends on the feedback. For instance, COMPCODER~\cite{compcode} uses the compiler feedback to repair code but misses run-time errors. D4C~\cite{xu2024aligning} aligns LLM output with their training objective for effective whole-program refinement without prior fault localization. LLM-CompDroid~\cite{liu2024llmcompdroid} enhances Android app reliability by integrating LLMs with traditional tools to detect and repair XML configuration compatibility bugs. RING~\cite{joshi2023repair} is a multilingual repair engine for correcting last-mile coding errors across multiple languages. ChatRepair~\cite{xia2023keep} uses a conversation-driven approach with prior test failure information to generate patches. Similarly, Conversational ACR~\cite{xia2023conversational} validates generated patches against a test suite, though test suite-based testing lacks completeness and may be inconsistently available.

\begin{table*}[ht!]
    \centering
    \setlength{\abovecaptionskip}{-5pt}
    \setlength{\belowcaptionskip}{-10pt}
    \caption{Comparison of related software bug detection and repair approaches.}
    \label{tab:Tool_comparison}
    \scriptsize
    \center
    \renewcommand{\arraystretch}{1.5}
    \rowcolors{2}{gray!25}{white}
    \begin{tabular}{|c|c|c|>{\centering\arraybackslash}p{3.5cm}|>{\centering\arraybackslash}p{2.2cm}|c|c|>{\centering\arraybackslash}p{2.8cm}|}
    \hline
    \multicolumn{5}{|c|}{\textbf{Framework details}} & \multicolumn{3}{c|}{\textbf{Repair}}\\ \hline
    \textbf{Name} & \textbf{Year} & \makecell{\textbf{Open} \\ \textbf{Source}} & \textbf{Dataset} & \textbf{Language} & \textbf{Granularity}& \textbf{Compiles} & \textbf{Method} \\ \hline\hline

    Bhayat et al.~\cite{bhayat2021towards} & 2021 & \ding{55} & SV-COMP~\cite{beyer2021sv-comp21} & C/C++ &N/A & N/A &N/A  \\ \hline
    OpenGBF~\cite{aljaafari2022combining} & 2022 & \cmark & SV-COMP~\cite{beyer2021sv-comp21} & C/C++ &N/A  & N/A & N/A \\ \hline
    ESBMC-Solidity~\cite{song2022esbmc} & 2022 & \cmark & Own\footnote{\url{https://github.com/esbmc/esbmc/tree/master/regression/esbmc-solidity}} & Solidity & N/A & N/A  & N/A \\ \hline
    FuseBMC \cite{alshmrany2022fusebmc} & 2022 & \cmark & Test-Comp~\cite{beyer2007path} & C/C++ & N/A & N/A  &N/A  \\ \hline
    COMPCODER~\cite{compcode} & 2022 & \ding{55} & AdVTest~\cite{lu2021codexglue}, CodeSearchNet~\cite{husain2019codesearchnet} & Python & Program & \cmark & Compiler Feedback based code completion \\ \hline
    Jigsaw~\cite{jain2022jigsaw} & 2022 & \ding{55} & PandasEval1, PandasEval2~\cite{jain2022jigsaw}\footnote{\url{https://github.com/microsoft/JigsawDataset/tree/main/datasets}} & Python & Snippets & \ding{55} & Program Synthesis \\ \hline
    Conversational ACR~\cite{xia2023conversational} & 2023 & \ding{55} & QuixBugs~\cite{lin2017quixbugs} & Java, Python & Function & \ding{55} & Prompt-based repair \\ \hline
    ChatRepair~\cite{xia2023keep} & 2023 & \ding{55} & Defects4J~\cite{just2014defects4j}, QuixBugs~\cite{lin2017quixbugs} & Java, Python & Patch & \ding{55} & Learns from previously failed tests \\ \hline
    Pearce et al.~\cite{pearce2022examining} & 2023 & \cmark & ExtractFix~\cite{gao2021beyond} & C, Python & Program & \cmark & Security tests-based \\ \hline
    RING~\cite{joshi2023repair} & 2023 & \ding{55} & BIFI~\cite{yasunaga2021break}, Bavishi et al.~\cite{bavishi2022neurosymbolic}, TFix~\cite{berabi2021tfix} & Excel,  C, PowerFx, PS, Python, JS & Program & \cmark & Compiler message \\ \hline
    Huang et al.~\cite{huang2023empirical} & 2023 & \cmark & Defects4J~\cite{just2014defects4j}, CPatMiner~\cite{li2022dear} & Java, C/C++, Python & Patch & \ding{55} & Model trained on buggy code - fix pair \\ \hline
    FuzzGPT~\cite{deng2023fuzzgpt} & 2024 & \ding{55} & Own~\cite{deng2023fuzzgpt} (unavailable) & Python & - & \ding{55} & LLM-based Fuzzing \\ \hline
    RepairAgent~\cite{bouzenia2024repairagent} & 2024 & \ding{55} & Defects4J~\cite{just2014defects4j} & Java & Program & \cmark & Invoking suitable tools \\ \hline
    SecRepair~\cite{islam2024llm} & 2024 & \ding{55} & InstructVul~\cite{islam2024llm} (unavailable) & C/C++ & Program & \cmark & Fine-tuned instruction training \\ \hline
    Self-Edit~\cite{zhang2023selfedit} & 2024 & \cmark & APPS~\cite{hendrycks2021measuring}, HumanEval~\cite{chen2021evaluating} & Python & Program & \cmark & Compile/Runtime with tests \\ \hline
    LLM-CompDroid~\cite{zhang2023selfedit} & 2024 & \ding{55} & ConfFix~\cite{huang2023conffix} & XML & Configuration & \ding{55} & Prompt-based \\ \hline
    ContrastRepair~\cite{kong2024contrastrepair} & 2024 & \ding{55} & Defects4J~\cite{just2014defects4j}, HumanEval~\cite{chen2021evaluating}, QuixBugs~\cite{lin2017quixbugs} & Java, Python & Program & \cmark & Contrastive test-pair \\ \hline
    CigaR~\cite{hidvegi2024cigar} & 2024 & \cmark & Defects4J~\cite{just2014defects4j}, HumanEval~\cite{chen2021evaluating} & Java & Patches & \ding{55} & Prompt optimization \\ \hline
    \textbf{ESBMC-AI} & \textbf{2024} & \cmark & FormAI~\cite{tihanyi2023formai} & C/C++ & Program & \cmark & Formal verification based feedback \\ \hline

    \end{tabular}
\end{table*}

Our work uses automated theorem provers to explore the uninvestigated combination of LLMs with FV techniques, particularly symbolic model checking. Table~\ref{tab:Tool_comparison} gives a quick view of how we position our \esbmcai framework concerning existing work. A desirable balance between two disparate concepts, symbolic verification and DL, can enhance the quality and speed of program repair. Relevant feedback that can be obtained from state-of-the-art software model checkers, such ESBMC~\cite{gadelha2018esbmc}, can show massive improvements in the patches suggested by GPTs.

\section{Background: Formal Verification Meets Large Language Models}
\label{sec:preliminaries}

BMC and LLMs are complementary techniques used in software engineering and artificial intelligence, respectively, and they are not directly connected. Given the current knowledge of automated reasoning and software verification, both methods have yet to be used to solve similar problems, such as software bug detection and debugging. Here, we use BMC to verify programs and provide diagnostic counterexamples via text to LLM. In contrast, LLM is used to understand the textual trace that leads to the program bug and thus tentatively produce code to fix the identified vulnerability.

\subsection{Bounded Model Checking (BMC)}
\label{bmc}

BMC is a primary component of our proposed counterexample-guided repair framework. State-of-the-art BMC engines support various industrial programming languages~\cite{CordeiroFM12, CordeiroKS19, SongMFC22, MonteiroGC22}. BMC represents the program as a state transition system extracted from the control-flow graph (CFG)~\cite{Aho:2006:CPT:1177220}, which is built during the translation from program text to Static Single Assignment (SSA) form. SSA is the ``language'' that the state-of-the-art SAT/SMT solvers understand, i.e., SSA expressions are converted to an SMT formula~\cite{CordeiroFM12}. A node in the CFG represents either a (non-) deterministic assignment, while an edge in the CFG represents a possible change in the program's control location.

We define a state transition system, denoted by $M$, as a triple $\left(S, R, s_1\right)$ where $S$ represents the set of states, $R \subseteq S \times S$ represents the set of transitions and $s_1 \subseteq S$ represents the set of initial states. A state $s \in S$ consists of the value of the program counter \textit{pc} and the values of all program variables. An initial state $s_{1}$ assigns the initial program location of the CFG to \textit{pc}. We identify each transition $T=(s_i,s_{i+1}) \in R$ between two states $s_{i}$ and $s_{i+1}$ with a logical formula $T(s_i,s_{i+1})$. This captures the constraints on the corresponding values of the program counter and the program variables.
 
We also define properties under verification in BMC: $\phi(s)$ is the logical formula encoding states satisfying a safety/security property, and $\psi(s)$ is the logical formula encoding states satisfying the completeness threshold, i.e., states corresponding to the program terminating. $\psi(s)$ will contain unwindings no deeper than the maximum number of loop iterations in the program. Note that, in our notation, termination, and error are mutually exclusive: $\phi(s) \wedge \psi(s)$ is by construction unsatisfiable; $s$ is a deadlock state if $T(s_i, s_{i+1}) \vee \phi(s)$ is unsatisfiable. The associated BMC problem is formulated by constructing the following logical formula:

\begin{equation}\label{eq:bmc}
  \text{BMC(k)} = I(s_1) \wedge \bigwedge^{k-1}_{i=1} T(s_i, s_{i+1}) \wedge
\bigvee^{k}_{i=1} \neg \phi(s_i).
\end{equation}

Here, $I$ the set of initial states of $M$ and $T(s_n, s_{n+1})$ is the transition relation of $M$ between time steps $i$ and $i+1$. Hence, $I(s_1)\wedge\bigwedge^{k-1}_{i=1} T(s_i, s_{i+1})$ represents the executions of $M$ of length $k$ and $BMC(k)$ can be satisfied if and only if for some $i \leq k$ there exists a reachable state at time step $i$ in which $\phi$ is violated. Suppose $BMC(k)$ is satisfiable. In that case, $\phi$ is violated, and the SMT solver provides a satisfying assignment from which we can extract the values of the program variables to construct a counterexample. 

We define a counterexample (or trace) for a violated property $\phi$ as a finite sequence of states $s_{1}, \ldots, s_{k}$ with $s_{1}, \ldots, s_{k} \in S$, and $T(s_i, s_{i+1})$ for $0 \leq i < k$. This sequence informs our LLM engine on reproducing the software vulnerability since it tells how to go from the program entry point to the property violation. Suppose that equation (\ref{eq:bmc}) is unsatisfiable. We could conclude that no error state is reachable in $k$ steps or less. In this case, we use this information to conclude that no software vulnerability exists in the program up to the bound $k$.

In our method, counterexamples are pivotal in guiding the LLM model. They enable the LLM to propose code corrections, taking inspiration from these counterexamples. Each counterexample specifies the exact trace, line number, and variable name, effectively highlighting the issue within the code. Without these counterexamples, even a simple code, as observed in the motivation section, could pose challenges for the LLM in suggesting a suitable fix. Further, it is essential to note that these counterexamples are based on rigorous mathematical proofs of whether a property holds. Consequently, the likelihood of introducing false positive findings is reduced to a very minimal level (though implementation errors may still exist), unlike results from simple static analysis tools. 

\subsection{Large Language Models (LLMs)}
\label{llm}

LLMs are DL systems based on the transformer architecture. They can understand, process, and generate human-like natural language. The input to an LLM consists of a sequence of tokens representing words, subwords, or characters transformed into a high-dimensional vector space using an embedding technique. These embedded tokens pass through multiple network layers, each applying non-linear transformations governed by learnable parameters. The output is often a probability distribution over possible next tokens, with the model selecting the highest probability token.
While LLMs are less efficient than state-of-the-art BMC tools for exact arithmetic operations and bounded model checking tasks, they excel in various natural language processing tasks, such as translation, question answering, and text generation. Transforming violated properties into human-like sentences enhances the LLM's understanding of code issues, allowing BMC counterexamples to correct erroneous code effectively.

Tom et al.~\cite{brown2020language} introduced GPT-3, the third iteration of the Generative Pretrained Transformer model developed by OpenAI. This paper's primary focus is on the few-shot learning capability of language models. The authors demonstrate that language models start exhibiting remarkable few-shot performance when scaled up, essentially learning from a limited number of examples. Lampinen et al.~\cite{lampinen2022can} investigated how AI systems interpret, understand, and apply knowledge from explanations provided in various contexts. Specifically, this is an important contribution to AI, particularly in language understanding and knowledge acquisition by machine learning models. Training or fine-tuning a transformer-based LLM, such as GPT-4~\cite{gpt4}, BERT~\cite{devlin2018bert}, T5~\cite{t5}, typically involves providing the model with a substantial volume of data in the form of input-output pairs. 

In this task, our inputs are the preprocessed counterexamples from BMC, and the outputs are human-readable interpretations of those counterexamples. When training an LLM, the model uses the ``Scaled Dot-Product Attention'' and ``Multi-Head Attention''~\cite{vaswani2017attention}. The attention mechanism allows the model to focus on different parts of the input sequence when producing the output sequence, which is especially useful for translating between complex BMC outputs and human language.  Mathematically, the scaled dot-product attention is calculated as:
\begin{equation}
\text{Attention}(Q, K, V ) = \text{softmax}\left(\frac{QK^T}{\sqrt{d_k}}\right)V,
\end{equation}

\noindent where $Q$, $K$, and $V$ are queries, keys, and values, respectively, and $d_k$ is the dimension of the queries and keys. This attention function is used in parallel or in ``heads'', enabling the model to focus on different features in the input. While scaled dot-product attention is frequently employed during training, it also proves to be highly valuable in the inference phase, showcasing a proficient understanding of BMC counterexamples.

Counterexamples provided by the BMC module often contain important but disconnected information, making them difficult for previous solutions, especially pre-transformer models, to interpret. The transformer's attention mechanism, specifically the scaled dot-product attention, enhances understanding complex inputs. For instance, consider the counterexample \texttt{"overflow line 7 function main, ERROR: argv[0]=32768, *mul(y,y)"}. Interpreting this requires the language model to understand multiple aspects. This counterexample shows an overflow in the variable $y$ during multiplication at line number 7.

The Scaled Dot-Product Attention can focus on different input parts based on their relevance to the current context. In this case, it could identify the link between the overflow error, the \texttt{mul(y,y)} function, and the specific line number mentioned. In other words, it can ``attend'' to the related information about the overflow error and the associated line of code when recommending an appropriate code fix.

This ability to dynamically allocate attention based on the input's content is one of the main reasons why transformer-based models such as GPT have succeeded across various tasks, including code debugging and automatic repair. They can understand the context of a given input, including intricate relations between separated segments, enabling them to suggest more accurate and relevant solutions or recommendations.

\section{Methodoology }
\label{sec:experiments}

\esbmcai is an AI-powered platform designed to expedite the detection and repair of critical software components. It employs a BMC tool in the background to identify vulnerabilities using formal verification methods such as abstract interpretation, constraint programming, and symbolic model checking, after which the generated counterexample is provided to the LLM with a specially crafted prompt. 

\subsection{Why ESBMC?}

Our ACR methodology can be implemented with various BMC tools. We chose the Efficient SMT-based Context-Bounded Model Checker (ESBMC)~\cite{gadelha2018esbmc} to implement our approach towards building self-healing software via LLMs and formal verification methods, illustrated in Figure~\ref{bmc-llm-approach}. In particular, we chose ESBMC since it is an efficient software verifier that can solve the highest amount of reachability-safety verification tasks within $10$ seconds time-limit according to SV-COMP 2024~\cite{beyer2024state}. We note that the selection of a 10-second time limit is not arbitrary. While increasing the time limit could yield improved results, longer processing times are unsuitable for code fixing in a live Integrated Development Environment (IDE) and continuous integration (CI) pipeline. By adhering to this limit, one can apply the proposed approach to such an existing framework and provide nearly real-time feedback to the programmer.

\subsection{User Chat Mode (UCM) and Fix Code Mode (FCM)}

\esbmcai currently operates in two distinct modes: the \emph{User Chat Mode (UCM)} and the \emph{Fix Code Mode (FCM)}. The primary purpose of the UCM mode is to utilize the capabilities of LLMs to simplify and clarify the complex counterexamples of ESBMC for the user. The huge amount of training data that LLMs have been trained with, along with their architecture, allows LLMs to use in-context learning of the counterexample; this allows the LLMs to generate high-quality explanations of a wide variety of problems and counterexamples. In UCM, users can pose questions to \esbmcai, such as \textit{``How can I correct this code?''} or \textit{``What is the problematic line of code?''}, among others. Based on the output from ESBMC, the LLM engine can offer valuable advice to the user, which may be implemented into the software according to the user's choice. In this mode, there is no automated code repair. Yet, the suggestions are grounded in the ESBMC output, which tends to be more precise than identifying specific bugs without formal verification methods.
 
Our primary focus is on FCM, aiming an advanced environment for identifying bugs and performing automatic code repair while ensuring the code remains compilable and retains its original behavior. In this mode, we utilize the well-recognized and industry-adopted ESBMC tool to detect vulnerabilities and leverage LLMs to fix the code. This presents challenges: we require a large and reliable dataset to evaluate our methodology, and human experts must carefully evaluate the applied patches to assess the success of the LLM in code rectification. Specialized prompts for each vulnerability are required to ``interpret'' the ESBMC counterexamples for an LLM. Human experts with a formal verification and software security background craft these prompts. For example, distinct prompts are required to address dereference failure versus buffer overflow in \texttt{scanf()}. Utilizing a general prompt such as ``fix the code based on this counterexample'' will significantly reduce accuracy in ACR.

\subsection{The ESBMC-AI Evaluation Dataset}

We need a sufficient number of vulnerable code samples to evaluate the effectiveness of the \esbmcai methodology. To showcase the strength of our methodology fully, we must note that not all datasets are suitable for our needs. The samples must be compilable, and the dataset should be labeled with the appropriate vulnerability class. Most available datasets~\cite{sard-black2018software, msr-cwe, zhou2019devign, Reveal-chakraborty2021deep} do not cater to at least one of these requirements.

The FormAI~\cite{tihanyi2023formai} dataset comprises $112,000$ AI-generated C programs, with $51.24$\% containing at least one vulnerability. The dataset covers diverse tasks, including complex ones like network management, encryption, table games, and simpler tasks like string manipulation. All C codes are compilable, and every C program in the dataset is labeled using a bounded model checking methodology with a $k=1$ bound parameter. Overall, we created $50,000$ samples for our evaluation. We then classified each program sample using ESBMC 7.6.1 and saved the results. To enhance vulnerability detection in each sample, we transitioned from bounded to unbounded model checking with unlimited \textit{k}-steps and a $500$-second timeout. This method strengthens the original approach applied in the FormAI dataset, where classification is based on bounded model checking with a $30$-second timeframe~\cite{tihanyi2023formai}. This process is very time-consuming and resource-intensive, even for small C programs. We used an Amazon AWS r7i.48xlarge instance with an AMD EPYC 9R14 CPU family featuring 192 vCPUs and 1.5TB of DDR5 RAM to handle this. Once the dataset was prepared and we identified which C samples were vulnerable and which were not, we applied our ESBMC-AI ACR methodology to attempt to fix the vulnerabilities. We randomly selected samples from eight popular vulnerability categories from the FormAI dataset (see Table~\ref{tab:recommendations}) for manual inspection and to verify the correctness of our approach. 

\section{Experimental Results}

This section presents the outcomes of integrating LLMs and BMC in \esbmcai, addressing the three research questions proposed at the beginning.

 We aim to answer these questions through our experiments and provide an in-depth statistical analysis of the results, offering comprehensive insights into the effectiveness of the \esbmcai approach and potential future improvements.

Let us denote all the $50000$ C samples by $\Sigma$, such that $\Sigma = \{c_1, c_2, \ldots, c_{50000}\}$, where each $c_i$ represents an individual sample. The samples can be divided into three primary categories: Verification Successful ($\mathcal{VS}$), Verification Failed ($\mathcal{VF}$), and Verification Unknown ($\mathcal{VU}$). These categories are mutually exclusive, meaning a single sample cannot belong to more than one category.  Our main focus is the $\mathcal{VF}$ category, which includes $31801$ samples, indicating that $63.60$\% of the code is vulnerable. The vulnerable samples can also be divided into three main subcategories: dereference failures ($\mathcal{DF}$), arithmetic overflow issues ($\mathcal{AO}$), and buffer overflow issues. The precise distribution of vulnerabilities in our dataset is shown in Table~\ref{tab:vulnerabilities}.

\begin{table}[htbp]
\centering
\caption{Top 32 Vulnerabilities in the 50000 dataset}
\begin{tabular}{cllr}
\toprule
\textbf{Cat} & \textbf{Violation Type} & \textbf{Count (\%)} \\
\midrule
\multicolumn{3}{c}{\cellcolor{gray!25}Vulnerability distribution} \\
\midrule
$\mathcal{DF}$ & Dereference failure: NULL pointer & \cellcolor{gray!30}14,700 (23.49\%) \\
$\mathcal{BO}$ & Buffer overflow on \texttt{scanf} & \cellcolor{gray!30}13,518 (21.60\%) \\
$\mathcal{DF}$ & Dereference failure: forgotten memory & \cellcolor{gray!30}7,681 (12.27\%) \\
$\mathcal{DF}$ & Dereference failure: invalid pointer & \cellcolor{gray!30}5,487 (8.77\%) \\
$\mathcal{DF}$ & Dereference failure: array bounds violated & \cellcolor{gray!30}4,020 (6.42\%) \\
$\mathcal{AO}$ & Arithmetic overflow on add & \cellcolor{gray!30}2,761 (4.41\%) \\
$\mathcal{AO}$ & Arithmetic overflow on sub & \cellcolor{gray!30}2,349 (3.75\%) \\
$\mathcal{DF}$ & Array bounds violated: upper bound & \cellcolor{gray!30}1,893 (3.02\%) \\
$\mathcal{DF}$ & Array bounds violated: lower bound & \cellcolor{gray!30}1,521 (2.43\%) \\
$\mathcal{AO}$ & Arithmetic overflow on mul & \cellcolor{gray!30}1,145 (1.83\%) \\
$\mathcal{DF}$ & DF: invalidated dynamic object & \cellcolor{gray!30}977 (1.56\%) \\
$\mathcal{BO}$ & Buffer overflow on \texttt{fscanf} & \cellcolor{gray!30}961 (1.54\%) \\
$\mathcal{AO}$ & Arithmetic overflow on FP ieee\_mul & \cellcolor{gray!30}943 (1.51\%) \\
$\mathcal{DF}$ & Division by zero & \cellcolor{gray!30}631 (1.01\%) \\
$\mathcal{AO}$ & Arithmetic overflow on FP ieee\_div & \cellcolor{gray!30}591 (0.94\%) \\
$\mathcal{DF}$ & VLA size overflows address space & \cellcolor{gray!30}507 (0.81\%) \\
$\mathcal{BO}$ & Buffer overflow on \texttt{sscanf} & \cellcolor{gray!30}498 (0.80\%) \\
$\mathcal{AO}$ & Arithmetic overflow on FP ieee\_add & \cellcolor{gray!30}497 (0.79\%) \\
$\mathcal{DF}$ & DF: Access to object OOB & \cellcolor{gray!30}453 (0.72\%) \\
$\mathcal{AO}$ & Arithmetic overflow on FP ieee\_sub & \cellcolor{gray!30}297 (0.47\%) \\
$\mathcal{DF}$ & File pointer must be valid & \cellcolor{gray!30}234 (0.37\%) \\
$\mathcal{DF}$ & DF: accessed expired variable pointer & \cellcolor{gray!30}199 (0.32\%) \\
$\mathcal{AO}$ & Arithmetic overflow on shl & \cellcolor{gray!30}170 (0.27\%) \\
$\mathcal{DF}$ & DF: write access to string constant & \cellcolor{gray!30}147 (0.23\%) \\
$\mathcal{AO}$ & Arithmetic overflow on div & \cellcolor{gray!30}137 (0.22\%) \\
$\mathcal{DF}$ & DF: incompatible base type & \cellcolor{gray!30}64 (0.10\%) \\
$\mathcal{DF}$ & DF of non-dynamic memory & \cellcolor{gray!30}60 (0.10\%) \\
$\mathcal{DF}$ & Free operand must have zero offset & \cellcolor{gray!30}44 (0.07\%) \\
$\mathcal{AO}$ & Arithmetic overflow on modulus & \cellcolor{gray!30}41 (0.07\%) \\
$\mathcal{DF}$ & Same object violation & \cellcolor{gray!30}34 (0.05\%) \\
$\mathcal{AO}$ & Arithmetic overflow on neg & \cellcolor{gray!30}18 (0.03\%) \\
$\mathcal{DF}$ & DF: Oversized field offset & \cellcolor{gray!30}7 (0.01\%) \\
\bottomrule
\end{tabular}
\label{tab:vulnerabilities}
\end{table}

Our primary objective is to fix as many programs as possible in each category. Our experiment used \texttt{GPT-4o} as the base LLM model within our ESBMC-AI framework. The formal verification tool ESBMC was invoked with the following flags in the background: \texttt{--overflow --memory-leak-check --show-stacktrace  --timeout 10 --unwind 1  --multi-property --no-unwinding-assertions --verbosity 6}. While the Form AI dataset used in this work contains C samples labeled with ESBMC, in a practical scenario, the vulnerability information is often unavailable for a given code. We employ a \textit{``real-time formal verification"} in the background to identify any vulnerabilities. If a vulnerability is found, the \esbmcai framework transforms the counterexample into the appropriate format and generates a corresponding prompt based on a previously created template by human experts. Given LLMs' sensitivity to prompts~\cite{cao2023prompt, white2023prompt}, we tested many different prompts to effectively incorporate the counterexample (stack traces) into our prompt alongside the original code.
 
\subsection{Experimental result on automated code repair}

Comparing the original and suggested code automatically is currently infeasible due to the undecidability of program equivalence~\cite{undecidable}. Verifying if the generated code is semantically equivalent to the original is a complex task that existing ACR tools overlook. Manually reviewing all $30,000$ samples for correctness is challenging, so we selected a smaller subset of $1,337$ samples from the most common categories for manual verification. 
This ensures that the fixes are consistent with the original programs. While most verifications are straightforward, some complex fixes require a more detailed review. To introduce automation, we have incorporated metrics useful for evaluating a patch's impact, such as changes in the number of lines of code (LOC) and alterations in cyclomatic complexity (CC). Significant deviations in these metrics can indicate that the patch was not successful.

We have categorized the most common vulnerabilities along with their associated CWE numbers. CWE numbers can indicate which vulnerabilities are most prevalent. Dereference failures, such as ``\textit{forgotten memory}" and ``\textit{NULL pointer}," can encompass various types of vulnerabilities. Assigning appropriate CWEs to these categories helps us determine the most frequent vulnerabilities in real-life projects. We aim to focus on fixing these CWEs with the highest possible accuracy.

\noindent \textbf{Buffer overflow on \texttt{scanf} and \texttt{fscanf}:}
Buffer overflows on \texttt{scanf()} and \texttt{fscanf()} are among the most common buffer overflow vulnerabilities in applications\footnote{\url{https://cwe.mitre.org/top25/archive/2023/2023_stubborn_weaknesses.html}}. For this type of vulnerability, a buffer overflow occurs when the \texttt{scanf/fscanf} function reads more data than the allocated buffer space, leading to an overwritten adjacent memory. This can cause unpredictable behavior, crashes, or other security vulnerabilities. The primary CWE number for \texttt{scanf()} and \texttt{fscanf()} is CWE-120. The related CWE numbers for \texttt{scanf()} include CWE-20, CWE-121, and CWE-122, which pertain to input validation issues, stack-based buffer overflow, and heap buffer overflow. The associated CWE numbers for \texttt{fscanf()} are CWE-129, CWE-131, and CWE-628, which involve incorrect calculation of buffer size and function calls with incorrectly specified arguments.

For \texttt{fscanf()}, we reviewed $175$ C sample code fixes, of which $160$ were successful, $8$ failed verification, and $7$ had unknown verification results, resulting in a $90.40$\% accuracy rate. We found that in $160$ samples where ESBMC indicated a successful patch, the patches were correct, compilable, and did not alter the original program behavior. Similarly, for \texttt{fscanf()}, we checked $241$ programs, where $220$ were successful, $13$ failed verification, and $8$ had unknown verification results, leading to a $91.29$\% accuracy rate. Cyclomatic complexity (CC) can be a good indicator of how complex a patch is. The average CC for the vulnerable programs using \texttt{fscanf()} is $4.61$, whereas, for the patched versions, it is $5.62$. This change of 1 CC aligns with expectations, as \texttt{fscanf()} I/O file issues are typically corrected with an if-then-else statement, which generally adds +1 to the CC.

\noindent \textbf{Dereference Failure: forgotten memory:} 
This issue contains many vulnerabilities associated with various CWEs, such as CWE-825, CWE-401, CWE-404, and CWE-459. Upon reviewing the formal verification output and the patches for this issue, we found that the same issue often emerges on a new line when a dereference failure is patched on a particular line. Therefore, dereference failure issues are widespread and can be problematic to pinpoint as a ``one-line'' problem. The output of the formal verification typically reveals a chain of errors that lead to a particular line, such as \texttt{strcpy}, \texttt{memcpy}, or other functions not part of the original code. These include files that are not part of the fixation prompt. Thus, a future improvement could be to add these, including the prompts with the original source code and stacktraces, to achieve better accuracy. From $187$ samples, \esbmcai achieved a $48.66$\% success rate. No external header or C files were needed to understand the issue within these patches.

\begin{table*}[ht]
\setlength{\abovecaptionskip}{-5pt}
\setlength{\belowcaptionskip}{-10pt}
\caption{Accuracy of fixation after one iteration for different types of vulnerabilities}
\label{tab:recommendations}

\center
\renewcommand{\arraystretch}{1.5}
\rowcolors{2}{gray!25}{white}
\begin{tabular}{lccc|cccccc}\hline
 \multicolumn{4}{c|}{\textbf{Original Programs}} & \multicolumn{6}{c}{\textbf{Patched Programs}} \\ \hline
  \makecell{ \textbf{Vulnerability} \\ \textbf{ Type}} & \makecell{\textbf{Sample}\\\textbf{size}}&
  \makecell{\textbf{Avg}\\\textbf{LOC}}&
  \makecell{\textbf{Avg}\\\textbf{CC}}& $\mathcal{VS}$& $\mathcal{VF}$&
  $\mathcal{VU}$&
  \makecell{\textbf{Avg}\\\textbf{CC}}&
  \textbf{Accuracy} & 
  \makecell{ \textbf{Patches}\\ \textbf{ Verified}}
  \\ \hline \hline
   Array bounds violation (lower bound)  & 182 &  79.56  & 6.72  & 174 &  4  & 4 & 8.35 & \cellcolor{green}95.60\%&  \ding{52} \\  \hline
   Buffer overflow on fscanf (I/O error)  & 241 &  74.95  & 4.61  &  220 &  13  & 8 & 5.62 &  \cellcolor{green!70}91.29\%&  \ding{52} \\  \hline
   Buffer overflow on scanf  & 175 &  78.92  & 6.91  &  160 &  8  & 7 & 8.30 &  \cellcolor{green!50}90.40\%&  \ding{52} \\  \hline
 Division by zero  & 133 &  73.52  & 3.77  &  115 &  8  & 10 & 4.42 &  \cellcolor{green!30}86.47\%&  \ding{52} \\  \hline

  Dereferene Failure: NULL pointer  & 229 &  78.05  & 5.44 &  184 &  40  & 5 & 7.70 &  \cellcolor{green!20}80.35\%&  \ding{52} \\  \hline

Arithmetic overflow on add  & 73 &  74.9  & 4.45  & 52 &  16  & 5 & 5.17 &  \cellcolor{orange!10}70.27\%&  \ding{52} \\  \hline
Dereference Failure: forgotten memory & 187 &  79.70  & 5.53  & 91 &  83  & 13 & 6.49 &  \cellcolor{orange!40}48.66\%&  \ding{52} \\  \hline
   
     Array bounds violation (upper bound) & 117 &  81.69  & 5.74  & 48 &  65  & 4 & 6.59 & \cellcolor{red!15}41.03\% &  \ding{52} \\  \hline

\end{tabular}
\vspace{-10pt}
\label{tab:verification_results}
\end{table*}

\noindent \textbf{Array bounds violations:}
Surprisingly, there was a significant difference in the accuracy of fixing array-bound violations. Upper-bound violations achieved an impressive $95.60$\% success rate, while lower-bound violations had a relatively low success rate of $41.03$\%. Upon careful review, we identified that lower-bound errors are easier to fix and do not require complex calculations by an LLM. These errors are usually associated with user input reading when null terminator characters are removed (See Listing~\ref{lis:ex1}).

\begin{listing}[t]

\caption{Array bound violation fix (lower bound)}
\label{lis:ex1}
\begin{tcolorbox}[title={ORIGINAL / FIXED code }, coltitle=white, colbacktitle=black, colback=gray!10]
\footnotesize
\begin{minted}[escapeinside=||]{bash}

ORIGINAL code:
        line[strlen(line) - 1] = '\0';

FIXED:
        size_t len = strlen(line);
        if (len > 0 && line[len - 1] == '\n')
        {
            line[len - 1] = '\0';
        }
\end{minted}
\end{tcolorbox}
\end{listing}

Contrary to expectations, when fixing upper bound violations, LLMs (including GPT-4, Gemini-Pro, and others) often try to correct the code by adding +1 to the variable. However, this approach usually fails to eliminate the bug. Simply increasing the upper bound by one still leaves the same issue with the buffer size, as shown by the formal verification output of the original and patched code in Listing~\ref{lis:ex2}. In the original code, we have \texttt{(signed long int)bytes\_received < \textbf{80}} upper bound violations, and in the fixation, we still have the same issue but with an increased value \texttt{(signed long int)bytes\_received < \textbf{81}}.

\begin{listing}[t]

\caption{Wrong fix: Array bounds violation (upper bound) }
\begin{tcolorbox}[title={ESBMC 7.6.1 model verification output}, coltitle=white, colbacktitle=black, colback=lightgray,before skip=-2.0mm]
\footnotesize
\begin{verbatim}
Violated property (ORIGINAL code):
  file falcon180b-10616_fixed.c line 56 column 13
  array bounds violated: array `buffer' upper bound
  (signed long int)bytes_received < 80
---------------------------------------------
Violated property (FIXED code):
  file falcon180b-10616.c line 57 column 13 
  array bounds violated: array `buffer' upper bound
  (signed long int)bytes_received < 81
\end{verbatim}
\end{tcolorbox}
\label{lis:ex2}

\end{listing}

\noindent \textbf{Division by zero:}
The division by zero vulnerability, identified by CWE-369 and associated with CWE-691 (Insufficient Control Flow Management), is quite common in applications. We manually verified a total of $133$ samples. Of these, $115$ fixations were successful, $8$ failed verification and $10$ had unknown verification results. This results in an accuracy rate of $86.47$\% for fixing division by zero vulnerabilities.

\noindent \textbf{Arithmetic overflow on add:} Here, we achieved a modest accuracy of $70.27$\%  from $73$ samples since fixing an addition overflow often introduces a new overflow. Consider the following interesting example:
\begin{equation}
    X= (A+B) \times 1000
\end{equation}
If the overflow on addition is patched correctly by handling variables $A$ and $B$, a new issue, such as a floating-point IEEE multiplication overflow, could emerge. Our methodology fixes one code issue at a time, as addressing multiple issues in a single iteration can reduce the model's accuracy due to biased attention, particularly with arithmetic overflows. Therefore, achieving higher accuracy often requires more than one iteration for most overflow issues. However, by fixing the arithmetic overflow in the first iteration and the floating-point IEEE multiplication overflow in the second iteration, an accuracy of 88\% can be achieved on the same samples.

Table~\ref{tab:verification_results} presents the overall verification results by category, ranked from highest to lowest.

\subsection{LLMs fixation without external help }

In the \esbmcai framework, a key component is supporting the LLMs with formal verification proof from external sources. This approach significantly enhances the accuracy of the fixes and guides the LLMs in the right direction. Without the exact counterexamples and stack traces, LLMs can fix the issues with approximately $31-37$\% accuracy, compared to $80$\% to $90$\% accuracy with ESBMC output. This demonstrates the effectiveness of our methodology and the external boost provided by formal verification. In certain cases, LLMs suggest that specific errors are present in C code, even though this may not be true. Consider the C code fragment illustrated on the left-hand side in Figure~\ref{fig:md5example}.

\begin{figure*}[ht]\centering

\begin{tcolorbox}[sidebyside, 
                  lefthand width=.42\textwidth,
                  righthand width=.4\textwidth,
                  sidebyside align=top seam,
                  colback=lightgray,
                  width=0.85\textwidth]

\begin{tcolorbox}[title={Vulnerable C code (misleading)}, coltitle=white, colbacktitle=black, colback=gray!10]
\footnotesize
\begin{minted}[escapeinside=||]{c}
#include <stdio.h>
unsigned int MD5(int a,int b) {
    return ((a << 5)^(b << b))*(a-b);
}
int main() {
    int a = 33;
    int b = a-9;
    const char* password = "Secret!";
    int result=MD5(a,b);
    printf("Result: %d\n", result);
    return 0;}
\end{minted}

\end{tcolorbox}

\tcblower

\begin{tcolorbox}[title={ESBMC Verification output}, coltitle=white, colbacktitle=black, colback=gray!10]
\footnotesize
\begin{verbatim}
Counterexample:

State 5 file  gpt661.c line 4 func MD5 
--------------------------------------
Violated property:
  file gpt661.c line 5 function MD5
  arithmetic overflow on mul
  !overflow("*", a << 5 ^ b 
  corresponding to << b, a - b)

VERIFICATION FAILED
\end{verbatim}
\end{tcolorbox}

\end{tcolorbox}
\setlength{\abovecaptionskip}{-1pt}
\setlength{\belowcaptionskip}{-15pt}
\caption{The actual vulnerability may be overlooked by an LLM when misleading function names are used.}
\label{fig:md5example}
\end{figure*}

The model generates various recommendations to resolve the problem, including removing the embedded secret password, questioning the validity of the MD5 function, and highlighting the insecurity of MD5. However, it failed to recognize the actual issue: an arithmetic overflow. Consequently, when the code is compiled, an overflow occurs, resulting in an incorrect outcome of ``\texttt{Result: -671079136}''. However, the \esbmcai framework correctly identifies and fixes the vulnerability, thanks to the formal verification counterexample, which guides the LLM in the right direction, as shown on the right-hand side of Figure~\ref{fig:md5example}. Without this guidance, even after $10$ attempts, the most advanced model still incorrectly identifies issues such as MD5 cryptographic problems or other errors in the code, which is not true in our case. The code does not use MD5 or include an embedded secret password. These examples demonstrate how LLMs can face challenges when accurately calculating arithmetic operations or identifying vulnerable code without external assistance.

\subsection{Threats to the Validity}
\label{subsec:threats-to-the-validity}

\esbmcai heavily relies on the language model's understanding of code semantics, which may not always align perfectly with the program's intended behavior. This can lead to the generation of repairs that, although syntactically valid, do not effectively address the underlying bugs or even introduce new issues. Such incorrect repairs can impact the overall accuracy and reliability of the framework's performance evaluation, potentially undermining its effectiveness in real-world scenarios. Moreover, since LLMs are off-the-shelf products prone to hallucinations and lack explainability, there is an added layer of uncertainty in the generated solutions. This highlights the critical need for incorporating mechanisms that enhance the interpretability and reliability of LLMs within the \esbmcai framework to ensure robust and trustworthy code repair in practice.

\section{Conclusions and Future Work}
\label{sec:conclusion}

This paper introduces a novel framework for automated code repair that leverages the power of Large Language Models and Bounded Model Checking techniques. Our proposed \esbmcai framework demonstrates significant advancements over existing works in the field by effectively utilizing feedback from the BMC engine to enhance program repair capabilities. Our evaluation of \esbmcai on randomly selected samples from the five most frequent vulnerabilities in the FormAI dataset reveals varying accuracy for fixing different vulnerabilities.

The results indicate that in a single iteration, over 90\% accuracy can be achieved for buffer overflow on scanf/fscanf and array bounds violations (lower bound). More than 80\% accuracy is attainable for division by zero and dereference failure: NULL pointer. These vulnerabilities cover the top 75\% of the most frequent CWEs. Array bounds violations (upper bound) and dereference failure: forgotten memory, are the most challenging issues to fix using this method as they often involve other vulnerabilities or external headers or C files.

Integrating LLMs and formal verification into automatic code repair is a promising future research direction. The power of these models, coupled with appropriate prompts and feedback mechanisms, enables more effective and intelligent code repair. However, addressing potential limitations and challenges is crucial, such as the need for massive computational resources and the possible introduction of unintended vulnerabilities or overfitting specific code patterns. We believe further advancements in this area will continue revolutionizing software development practices by enabling faster and more accurate bug fixes, ultimately enhancing software reliability, productivity, and security.

We have released our tool and methodology on our project webpage, and \esbmcai 0.5.1 is now available as a PyPI module.This makes \esbmcai one of the few tools that can effectively support real-world projects, harnessing the full power of formal verification methods.

\bibliographystyle{IEEEtran}
\bibliography{reference}

\end{document}